\documentstyle[prl,aps,epsf,epsfig]{revtex} 
\begin{document}
\draft
\twocolumn[\hsize\textwidth\columnwidth\hsize\csname@twocolumnfalse\endcsname
%
\title{
Magnetic skyrmion lattices in heavy fermion superconductor UPt$_3$}
\author
{A. Knigavko$^1$ and B. Rosenstein$^2$}
\address{
$^1$Institute of Physics, Academia Sinica, Taipei, Taiwan 11529, R.O.C.\\
$^2$Electrophysics Department, National Chiao Tung University, Hsinchu, 
Taiwan 30043, R.O.C.
}

\maketitle

\begin{abstract}
Topological analysis of nearly SO(3)$_{spin}$ symmetric Ginzburg--Landau
theory, proposed for UPt$_{3}$ by Machida {\it et al}, shows that there
exists a new class of solutions carrying two units of magnetic flux: the
magnetic skyrmion. These solutions do not have singular core like Abrikosov
vortices and at low magnetic fields they become lighter for strongly type II
superconductors. Magnetic skyrmions repel each other as $1/r$ at distances
much larger then the magnetic penetration depth $\lambda ,$ forming a
relatively robust triangular lattice. The magnetic induction near $H_{c1}$
is found to increase as $(H-H_{c1})^{2}$. This behavior agrees well with
experiments.
\end{abstract}

]

\vspace{1cm}


Heavy fermion superconductors have surprising properties on both the
microscopic and the macroscopic level. Charge carrier's pairing mechanism is
unconventional. The vortex state is also rather different from that of
s-wave superconductors: there exist unusual asymmetric vortices, and phase
transitions between numerous vortex lattices take place\cite{Sauls1}. For
the best studied material UPt$_{3}$ several phenomenological theories have
been put forward \cite{Machida1,Sauls2,Joynt} which utilize a multicomponent
order parameter. In particular, great effort has been made to qualitatively
and quantitatively map the intricate $H-T$ phase diagram. Most attention has
been devoted to the region of magnetic fields near $H_{c2}$.

Magnetization curves of \ UPt$_{3}$ near $H_{c1}$ are also rather unusual
(see Fig.1). Theoretically, if the magnetization is due to penetration of
vortices into a superconducting sample then one expects $-4\pi M$ to drop
with an infinite derivative at $H_{c1}$ (dotted line). On the other hand
experimentally $-4\pi M$ continues to increase smoothly (squares and
triangles represent rescaled data taken from refs. \cite{Maple} and \cite
{Zhao} respectively). Such a behavior was attributed to strong flux pinning
or surface effects \cite{Maple}. However both experimental curves in Fig.1,
as well as the other ones found in literature, are close to each other if
plotted in units of $H_{c1}$. There might be a more fundamental explanation
of the universal smooth magnetization curve near $H_{c1}$. If one assumes
that fluxons are of unconventional type for which interaction is long range
\ then precisely this type of magnetization curve is obtained.

Magnetization near $H_{c1}$ due to fluxons carrying $N$ units of flux $\Phi
_{0}\equiv hc/2e$, with line energy $\varepsilon $ and mutual interaction $%
V(r)$, is found by minimizing the Gibbs energy of a very sparse triangular
lattice: 
\begin{equation}
G(B)=\frac{B}{N\Phi _{0}}\left[ \varepsilon +3V(a)\right] -\frac{BH}{4\pi },
\label{Gibbs-energy}
\end{equation}
where $a=(\Phi _{0}/B\sqrt{3})^{\frac{1}{2}}$ is lattice spacing. When $%
V(r)\sim \exp [-\lambda r],$ the magnetic induction has the conventional
behavior $B\sim \left[ \log \left( H-H_{c1}\right) \right] ^{-2}$\cite
{Tinkham}, while if it is long range, $V(r)\sim 1/r^{n},$ then one finds $%
B\sim \left( H-H_{c1}\right) ^{n+1}$. The physical reason for this different
behavior is very clear. For a short range repulsion, if one fluxon
penetrated the sample, many more can penetrate almost with no additional
cost of energy. This leads to the infinite derivative of magnetization. On
the other hand for a long range interaction making a place for each
additional fluxon becomes energy consuming. Derivative of magnetization thus
becomes finite.

It is generally assumed that although vortices in UPt$_{3}$ differ from the
usual Abrikosov vortices in many details \cite{Tokuyasu1} two important
characteristics are preserved. First, their size $\lambda $ is well defined:
magnetic field and interactions between vortices vanish exponentially beyond
this length. Second, their energy is proportional to $\log \kappa $.
However, in this note we show on the basis of topological analysis of a
model by Machida {\it et al. }\cite{Machida2} that there exists an
additional class of fluxons which we call magnetic skyrmions. They carry two
units of magnetic flux $N=2$ and do not have a singular core, similar to ATC
texture in superfluid $^{3}$He \cite{Salomaa}. We show that their line
energy $\varepsilon \approx 2\varepsilon _{0}$, $\varepsilon _{0}\equiv
\left( \Phi _{0}/4\pi \lambda \right) ^{2}$ is independent $\kappa $ and is
smaller than that of Abrikosov vortices for strongly type II superconductors
like UPt$_{3}$ ($\kappa \sim 50$). Magnetic skyrmion lattice becomes the
ground state at low magnetic fields $(H-H_{c1})/H_{c1}\ll 1$. We further
find that the repulsion of magnetic skyrmions is, in fact, long range: $%
V(r)\sim 1/r$. This allows us to produce a nice fit to the magnetization
curve (solid line in Fig.1) which is universal (independent of $\kappa $).

The order parameter in the weak spin--orbit coupling model of UPt$_{3}$ is a
three dimensional complex vector: $\psi _{i}(\vec{r})$\cite{Machida1}. The
Ginzburg--Landau free energy reads: 
\begin{equation}
F=F_{sym}+\Delta F,  \label{Tot-energy}
\end{equation}
\begin{eqnarray}
&F_{sym}=-\alpha \psi _{i}\psi _{i}^{\ast }+\frac{\beta _{1}}{2}(\psi
_{i}\psi _{i}^{\ast })^{2}+\frac{\beta _{2}}{2}|\psi _{i}\psi _{i}|^{2}
\label{Pot-energy} \\
&+K_{1}\left( |{\cal D}_{x}\psi _{i}|^{2}+|{\cal D}_{y}\psi
_{i}|^{2}\right) +K_{2}\left| {\cal D}_{z}\psi _{i}\right| ^{2}+\frac{1}{%
8\pi }B_{j}^{2},  \label{Grad-energy} \\
&\Delta F=-\gamma |\psi _{x}|^{2}-\lambda |\psi _{z}|^{2}+\frac{\Delta \chi 
}{2}|\psi _{i}B_{i}|^{2}  \label{Sym-break-pot} \\
&+\sum_{i=x,y}\left[ k_{1}^{i}\left( |{\cal D}_{x}\psi _{i}|^{2}+|{\cal D}%
_{y}\psi _{i}|^{2}\right) +k_{2}^{i}\left| {\cal D}_{z}\psi _{i}\right| ^{2}%
\right] ,  \label{Sym-break-grad}
\end{eqnarray}
where ${\cal D}_{j}\equiv \partial _{j}-i(2{e}/{\hbar c})A_{j}$ and $%
B_{j}=(\nabla \times \vec{A})_{j}$. We separated eq.(\ref{Tot-energy}) into
a symmetric part $F_{sym}$ which is invariant under the spin rotation group $%
\ $SO(3)$_{spin}$\ acting on the index $i$ of the order parameter, and into
terms breaking the SO(3)$_{spin}$ symmetry (anisotropy, coupling to
antiferromagnetic spin fluctuations and spin-orbit coupling). Although $%
\Delta F$ is crucial in explaining the double superconducting phase
transition in UPt$_{3}$ at zero external magnetic field and the shape of $%
H_{c2}(T)$ curve on the $H-T$ phase diagram, it can be considered as a small
perturbation in the low temperature superconducting phase (phase B) well
below its critical temperature $T\ll T_{c}^{-}\simeq .45$ K and at low
magnetic fields $H\simeq H_{c1}$. Indeed, estimation of coefficients of $%
\Delta F$\ at $T=T_{c}^{-}/2$ yields: $\gamma /\alpha \simeq .2,\lambda
/\alpha \simeq .05$ and $\left( \frac{\Delta \chi }{2}H_{c1}^{2}\right)
/\left( \frac{\alpha ^{2}}{2\beta _{1}}\right) \simeq 10^{-6}$, and also $%
k\ll K$ \cite{Machida1}. Therefore, in a certain range of magnetic fields
and temperatures there is an approximate O(3) symmetry and we first turn to
minimize $F_{sym}$.

In the vacuum of phase B the order parameter is $\vec{\psi}=\psi _{0}(\vec{n}%
+i\vec{m})/\sqrt{2}$,$\psi _{0}^{2}\equiv \alpha /\beta _{1}$, $\;\vec{n}%
\perp \vec{m}$, $\vec{n}^{2}=\vec{m}^{2}=1.$ Stability requires: $\alpha >0$%
, $\beta _{1}>0$ and $\beta _{2}>-\beta _{1}.$ The symmetry breaking pattern
of phase B is as follows. Both the spin rotations SO(3)$_{spin}$ and the
U(1) gauge symmetries are spontaneously broken, but a diagonal subgroup U(1)
survives. The subgroup consists of combined transformations: rotations by
angle $\vartheta $ around the axis\ $\vec{l}$ $\equiv \vec{n}\times \vec{m}$
accompanied by gauge transformations $e^{i\vartheta }$. Each vacuum state is
specified by orientation of a triad of orthonormal vectors $\vec{n},$ $\vec{m%
},$ $\vec{l}$. The vacuum manifold is therefore isomorphic to SO(3).
Topological defects might be of two kinds: regular and ''singular''. To find
regular topological line defects, it is enough to consider the London
approximation \cite{Salomaa}, i.e. to assume that the order parameter
gradually changes in space from one vacuum to another. The triad $\vec{n},$ $%
\vec{m},$ $\vec{l}$ then becomes a field. The Abrikosov vortex is a singular
defect: it has a core where the modulus of the order parameter vanishes and
energy diverges logarithmically. Accordingly, a cutoff parameter -- the
correlation length -- should be introduced and one obtains $\log \kappa $
dependence for the vortex line energy. This fact alone means that if there
exists a regular solution it is bound to become energetically favorable for
large enough $\kappa $. Below we consider a situation when the external
magnetic field is oriented along $z$ axis and all configurations are
translationally invariant in this direction. We use dimensionless units: $%
r\equiv \lambda \tilde{r}$, $A\equiv \left( \Phi _{0}/2\pi \lambda \right) 
\tilde{A}$, $B\equiv \left( \Phi _{0}/2\pi \lambda ^{2}\right) \tilde{B}$
and $F=(\varepsilon _{0}/2\pi \lambda ^{2})\tilde{F},$ where $\lambda \equiv
(\Phi _{0}/2\pi )\sqrt{\beta _{1}/4\pi \alpha K_{1}}$ (the tilde will be
dropped hereafter). The free energy takes form 
\begin{equation}
F_{L}=1/2\left( \partial _{k}\vec{l}\right) ^{2}+\left( \vec{n}\partial _{k}%
\vec{m}-A_{k}\right) ^{2}+B_{k}^{2}  \label{main functional}
\end{equation}
and the field equations are 
\begin{eqnarray}
&n_{p}\vec{\nabla}m_{p}\!\!-\vec{A} =\vec{\nabla}\times \left( \vec{\nabla}%
\times \vec{A}\right) =\vec{j},  \label{current-eq} \\
&\Delta \vec{l}-\vec{l}(\vec{l}\cdot \Delta \vec{l})+2j_{k}(\vec{l}\times
\partial _{k}\vec{l}) =0.  \label{OP-eq}
\end{eqnarray}
Eq.(\ref{current-eq}) shows that the superconducting velocity is given by $%
n_{p}\vec{\nabla}m_{p}=-\vec{\nabla}\vartheta $, where the angle $\vartheta $
specifies the orientation of vector $\vec{n}$ or $\vec{m}$ in the plane
perpendicular to $\vec{l}$ (see insert in Fig.2). Thus, $\vartheta $ is the
superconducting phase.

Now we proceed to classify the boundary conditions. Magnetic field vanishes
at infinity, while topology of the orientation of the triad $\vec{n},$ $\vec{%
m},$ $\vec{l}$ at different distant points is described by the first homotopy 
group of vacuum manifold: $\pi _{1}$(SO(3))$=$Z$_{2}$ \cite{Salomaa}. 
It yields a classification of solutions into two topologically distinct
classes (''odd'' and ''even''). This classification is too weak, however,
for our purposes because it does not guarantee nontrivial flux penetrating
the plane. We will see that configurations having both ''parities'' are of
interest. In the presence of the magnetic flux possible configurations are
further constrained by the flux quantization condition. The vacuum manifold
is naturally factored into SO(3)$\rightarrow $SO(2)$\otimes $S$_{2}$
where the S$_{2}$ is set of directions of $\vec{l}$ and the SO(2) is the
superconducting phase $\vartheta $. For a given number of flux quanta $N$,
the phase $\vartheta $ makes $N$ windings at infinity, see Fig. 2. The first
homotopy group of this part is therefore fixed: $\pi _{1}$(SO(2))$=$Z$.$ If,
in addition, $\vec{l}$ is constant, there is no way to avoid singularity in
the phase $\vartheta $ where $|\vec{\psi}|=0$. However, general requirement
that a solution has finite energy is much weaker. It tells us that the
direction of $\vec{l}$ should be fixed only at infinity \cite{footnote}. The
relevant homotopy group is nontrivial: $\pi _{2}$(S$_{2} $)$=$Z. The second
homotopy group appears because fixing $\vec{l}$ at infinity (say, up)
effectively ''compactifies'' two dimensional physical space into S$_{2}.$
Unit vector $\vec{l}$ winds towards the center of the texture. The new
topological number is $Q=(1/8\pi )\int \varepsilon _{ij}\,\vec{l}\,\left(
\partial _{i}\vec{l}\times \partial _{j}\vec{l}\right) d^{2}r $. \
Therefore, all configurations fall into classes characterized by the two
integers $N$ and $Q$. For regular solutions, however, these two numbers are
not independent. Upon integrating the supercurrent equation, eq.(\ref
{current-eq}) along a remote contour and making use of the identity $%
\varepsilon _{pqs}l_{p}(\partial _{i}l_{q})(\partial _{j}l_{s})=(\partial
_{i}n_{p})(\partial _{j}m_{p})-(\partial _{i}m_{p})(\partial _{j}n_{p}),$ we
obtain: $Q=N/2.$ We call these regular solutions magnetic skyrmions.

The lowest energy solution within the London approximation corresponds to $%
N/2=Q=-1$ (or $N/2=Q=+1$). We analyze a cylindrically symmetric situation
and choose the triad $\vec{n},$ $\vec{m},$ $\vec{l}$ in the form: 
\begin{eqnarray}
\vec{l} &=&\vec{e}_{z}\cos \Theta (\rho )+\vec{e}_{\rho }\sin \Theta (\rho ),
\nonumber \\
\vec{n} &=&\left( \vec{e}_{z}\sin \Theta (\rho )-\vec{e}_{\rho }\cos \Theta
(\rho )\right) \sin \varphi +\vec{e}_{\varphi }\cos \varphi ,  \label{anzatz}
\\
\vec{m} &=&\left( \vec{e}_{z}\sin \Theta (\rho )-\vec{e}_{\rho }\cos \Theta
(\rho )\right) \,\cos \varphi -\vec{e}_{\varphi }\sin \varphi ,  \nonumber
\end{eqnarray}
where $\rho $ and $\varphi $ are polar coordinates and $\Theta =\widehat{%
\vec{e}_{z}\vec{l}}$. Boundary conditions are: $\Theta (0)=\pi $ and $\Theta
(\infty )=0$. The vector potential is given by $\vec{A}=A(\rho )\vec{e}%
_{\varphi }$. The general form of such a configuration is shown in Fig.2.
The unit vector $\vec{l}$ (solid arrows) flips its direction from up to down
as it moves from infinity toward the origin. The phase $\vartheta $ (arrow
inside small circles in Fig.2) winds twice while completing an ''infinitely
remote'' circle. If in eq. (\ref{main functional}) only the first term were
present we would deal with a standard $SO(3)$ invariant nonlinear $\sigma $%
-model \cite{Rajaraman}. Being scale invariant, it possesses infinitely many
pure skyrmion solutions $\Theta _{s}(\rho ;\delta )=2\arctan (\delta /\rho ),
$ which have the same energy equal to $2$ (in units of $\varepsilon _{0}$)
for any size $\delta $ of a skyrmion. However, in the present case the
structure of the order parameter is more complex and the above degeneracy is
lifted \ by the second and third terms of eq. (\ref{main functional}).

Below we make use of the functions $\Theta _{s}(\rho ;\delta )$ to
explicitly construct the variational configurations. We show that as size of
these configurations increases the energy is reduced to a value arbitrarily
close to the absolute minimum of $\varepsilon _{ms}=2$. Substituting eq.(\ref
{anzatz}) into eq.(\ref{main functional}) and integrating over the $x-y$
plane we obtain the energy of the magnetic skyrmion in the form: $%
\varepsilon _{ms}=\varepsilon _{s}+\varepsilon _{cur}+\varepsilon _{mag}$,
where $\varepsilon _{s}\equiv \int \rho d\rho \left( \Theta ^{\prime
2}/2+\sin ^{2}\Theta /2\rho ^{2}\right) $, $\varepsilon _{cur}\equiv \int
\rho d\rho \left[ (1+\cos \Theta )/\rho +A\right] ^{2}$ and $\varepsilon
_{mag}\equiv \int \rho d\rho \left( A/\rho +A^{\prime }\right) ^{2}$. The
first term $\varepsilon _{s}$ is the same as in  nonlinear $\sigma $-model
without magnetic field. It is bound from below by $2$, the energy of a pure
skyrmion. The second term $\varepsilon _{cur}$, the ''supercurrent''
contribution, is positive definite. One still can maintain zero value of
this term when the field $\Theta (\rho )$ is a pure skyrmion $\Theta
_{s}(\rho ;\delta )$ of certain size $\delta $. Assuming this one gets: $%
A(\rho )=-(1+\cos \Theta )/\rho =-2\rho /(\rho ^{2}+\delta ^{2})$. The third
term, the magnetic field contribution (which is also positive definite),
becomes $\varepsilon _{mag}=8/3\delta ^{2}$. It is clear that when $\delta
\rightarrow \infty $ we obtain energy arbitrarily close to the lower bound: $%
\varepsilon _{ms}\leq 2+8/3\delta ^{2}\rightarrow 2$. Single magnetic
skyrmion therefore blows up.

If many magnetic skyrmions are present, then their interactions can
stabilize the system. They repel each other, as we will see shortly, and
therefore form a lattice. Since they are axially symmetric, the interaction
is axially symmetric and thus a triangular lattice is expected. Assume that
the lattice spacing is $a$. At the boundaries of the hexagonal unit cells
the angle $\Theta $ is zero, while at the centers it is $\pi $. The magnetic
field $B$ is continuous on the boundaries. Therefore, to analyze a magnetic
skyrmion lattice we should solve eqs.(\ref{current-eq})--(\ref{OP-eq}) on
the unit cell with these boundary conditions demanding that two units of
flux pass through the cell (by adjusting the value of magnetic field on the
boundary). We have approximated the hexagonal unit cell by a circle of
radius $R=3^{3/4}a/\sqrt{\pi }$ and the same area, and performed numerical
integration of the equations

\begin{eqnarray}
A^{\prime \prime }+\frac{A^{\prime }}{\rho }-\frac{A}{\rho ^{2}}-A-\frac{%
1+\cos \Theta }{\rho } &=&0,  \label{rot-eq1} \\
\Theta ^{\prime \prime }+\frac{\Theta ^{\prime }}{\rho }+\frac{\sin \Theta }{%
\rho }\left( \frac{2+\cos \Theta }{\rho }+2A\right)  &=&0,  \label{rot-eq2}
\end{eqnarray}
which follow from the cylindrically symmetric Ansatz of eq.(\ref{anzatz}).
Calculations for $R$ from $R=5$ till $R=600$ were done by means of a finite
element method. The energy per unit cell in a wide range of $R$ is
satisfactory described (deviation at $R=10$ is $1\%$) by the function $%
\varepsilon _{cell}\simeq 2+5.62/R$. Note that in the limit $R\rightarrow
\infty $ we recover our previous variational estimate: $\varepsilon
_{cell}\rightarrow \varepsilon _{ms}=2.$ The dominant contribution to
magnetic skyrmion energy at large $\ R$ comes from the first term $%
\varepsilon _{s}$, similar to the analytical variational state described
above. The contribution to $\varepsilon _{cell}$ from magnetic field, $%
\varepsilon _{mag}$, is small for large $R$ but becomes significant in
denser lattices. The most interesting feature of the solution is that the
supercurrent contribution $\varepsilon _{cur}$ to the energy of magnetic
skyrmion is negligibly small for all considered values of $R$. This is to be
compared with the usual Abrikosov vortex where at high $\kappa $ the total
energy is dominated by magnetic and supercurrent contributions which are of
the same order of magnitude. Most of the flux goes through the region where
the vector $\vec{l}$ is oriented upwards. In other words, the magnetic field
is concentrated close to the center of a magnetic skyrmion.

Line energy of Abrikosov vortices $\varepsilon _{v}$ for the present model
was calculated numerically (beyond London approximation) in \cite{our}. For $%
\kappa =20$ and \ $50$ we obtain $2\varepsilon _{v}/\varepsilon _{ms}\approx
3.5$ and $4.4$ respectively. Therefore we expect that the lower critical
field of UPt$_{3}$ is determined by magnetic skyrmions: $h_{c1}=\varepsilon
_{ms}/2N$. Returning to physical units, 
\begin{equation}
H_{c1}=\Phi _{0}/4\pi \lambda ^{2}.  \label{Hc1}
\end{equation}
To find magnetization, we now utilize eq.(\ref{Gibbs-energy}). Interactions
among magnetic skyrmions follows easily from the energy of a unit cell of
the hexagonal lattice: $V(r)=2(\varepsilon _{cell}-2)/6\simeq 1.87/r$. The
resulting averaged magnetic induction, in units of $\Phi _{0}/2\pi \lambda
^{2}$, reads 
\begin{equation}
B\simeq 0.25\left( H/H_{c1}-1\right) ^{2}.\;\;  \label{magnetization}
\end{equation}
This agrees very well with the experimental results, see Fig.1. For fields
higher then several $H_{c1}$ London approximation is not valid anymore since
magnetic skyrmions will start to overlap. In this case one expects that
ordinary Abrikosov vortices, which carry one unit of magnetic flux, become
energetically favorable. The usual vortex picture has indeed been observed
at high fields by Yaron{\it \ et al}. \cite{Yaron} Curiously, our result is
similar to conclusions of Burlachkov {\it et al.}\cite{Burlachkov} who
investigated stripe-like (quasi one dimensional) spin textures in triplet
superconductors. Having established the magnetic skyrmion solution of $%
F_{sym}$ we next estimated how it is influenced by various terms of $\Delta F
$ eqs.(\ref{Sym-break-pot})--(\ref{Sym-break-grad}). It was found that these
perturbations do not lead to destabilization of a magnetic skyrmion.

In conclusion, we have performed a topological classification of the
solutions in SO(3)$_{spin}$ symmetric GL free energy. This model, with
addition of very small symmetry breaking terms, describes heavy fermion
superconductor UPt$_{3}$ and possibly other p-wave superconductors. A new
class of topological solutions in weak magnetic field was identified. These
solutions, magnetic skyrmions, do not have normal core. At small magnetic
fields the magnetic skyrmions are lighter then Abrikosov vortices and
therefore dominate the physics. Magnetic skyrmions repel each other as $1/r$
at distances much larger then magnetic penetration depth forming a
relatively robust triangular lattice. $H_{c1}$ is reduced by a factor $\log
\kappa $ as compared to that determined by usual Abrikosov vortex (see eq.(%
\ref{Hc1})).

The following characteristic features, in addition to the slope of the
magnetization curve, can allow experimental identification of a magnetic
skyrmions lattice.

1. Unit of flux quantization is $2\Phi _{0}$.

2. Superfluid density $|\vec{\psi}|^{2}$ is almost constant throughout the
mixed state. This can be tested using STM techniques.

3. Due to the fact that there is no normal core, in which usually
dissipation and pinning take place, one expects that pinning effects are
reduced.

It is interesting to note that our results are actually applicable to
another model of UPt$_{3}$ with accidentally degenerate AE representations 
\cite{Zhitom1}. Although this model adopts the strong spin--orbit coupling
scheme, \ it has a structure closely related to $F_{sym}$ of eqs.(\ref
{Pot-energy})--(\ref{Grad-energy}) at low temperatures where both order
parameters become of equal importance and can be viewed as a single three
dimensional order parameter.

The authors are grateful to B. Maple for discussion of results of Ref. 5, to
L. Bulaevskii, T.K. Lee and J. Sauls for discussions and to A. Balatsky for
hospitality in Los Alamos. The work is supported by NSC, Republic of China,
through contract \#NSC86-2112-M009-034T.





\begin{figure}[htp]
\epsfig{figure=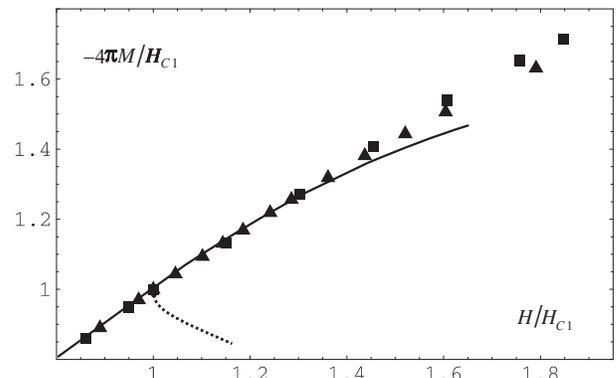,height=5cm,width=8cm,angle=-0}
\caption{Magnetization of magnetic skyrmion lattice (solid line) and
experimental data for UPt$_{3}$ from Ref.5 (squares) and Ref.6 (triangules).
Magnetization of Abrikosov vortex lattice (dotted line) is given for
comparison.}
\end{figure}


\begin{figure}[htp]
\epsfig{figure=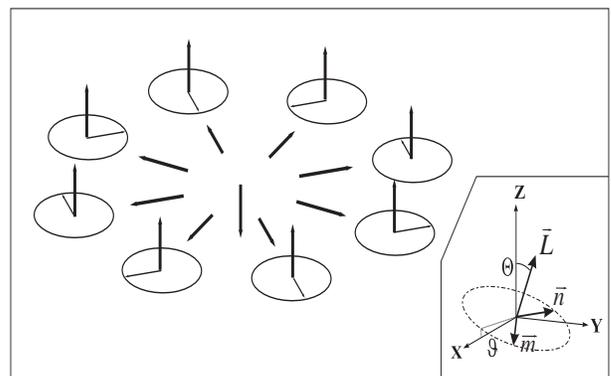,height=5cm,width=8cm,angle=-0}
\caption{ Configuration of a magnetic skyrmion. Solid arrows represent  
$\vec{l}$ field while the ''clocks'' show that phase $\protect\vartheta $
rotates twice as an infitely remote contour is circumvented. Insert explains
defintions of $\protect\vartheta $ and $\Theta .$ }
\end{figure}


\begin{references}
\bibitem{footnote}  This follows from the presence of $(\partial _{i}\vec{l}%
)^{2}$ term in $F_{sym}$ of eq. (\ref{main functional}), which cannot be
''gauged away'' similar to the corresponding term for the SO(2) part.
\end{references}
\end{document}